\documentclass[12pt]{article}
\usepackage{amssymb}
\usepackage{amsmath}
\usepackage{epsfig}
\usepackage{float}
\newcommand{\be}{\begin{equation}}
\newcommand{\ee}{\end{equation}}
\newcommand{\bea}{\begin{eqnarray}}
\newcommand{\eea}{\end{eqnarray}}
\begin{document}

\begin{center}
{\bf NEUTRINO  MAJORANA}

\end{center}

\begin{center}
S. M. Bilenky
\end{center}

\begin{center}
{\em  Joint Institute
for Nuclear Research, Dubna, R-141980, Russia, and\\
SISSA,via Beirut 2-4, I-34014 Trieste, Italy.}
\end{center}
\begin{abstract}
The Majorana paper ``Symmetrical theory of the electron and
positron'' is briefly reviewed. The present status of Majorana
neutrinos is discussed.
\end{abstract}

\section{Introduction}
This year we celebrate the 100-th  anniversary of the birth of
Ettore Majorana,  one of the greatest physicist of the XX  century. E.
Majorana was a very critical person.  Especially in the last years
of his short life, he  did not like to  publish his results. He
published his most important paper ``Symmetrical theory of the
electron and positron'' \cite{Majorana}  in 1937, probably, because of the
competition for the chair in theoretical physics at the Palermo
University (see \cite{Pont}). Fermi, Amaldi and other participants
of the Fermi group convinced Majorana to take part in the
competition. His previous paper (on nuclear forces) was published in
1933 when he was   visiting   Germany (Heisenberg convinced him at
that time to publish the paper).

I will discuss  here briefly   the content of the Majorana paper.
Majorana was not satisfied with the existing at that time theory of electrons and
positrons in which positrons were  considered as holes in the Dirac sea
of the states of electrons with negative energies. He wanted to
formulate the symmetrical theory in which there is no notion of
states with negative energies.

Let us consider a complex spinor field $\psi(x)$ which satisfies  the Dirac equation 
\begin{equation}\label{1a}
(i \gamma^{\alpha}\partial_{\alpha} -m)\,\psi(x)=0.
\end{equation}
The conjugated field
\begin{equation}\label{2a}
\psi^{c}(x)=C \bar\psi^{T}(x)
\end{equation}
obviously also satisfies the Dirac equation
\begin{equation}\label{3a}
(i \gamma^{\alpha}\partial_{\alpha} -m)\,\psi^{c}(x)=0.
\end{equation}
Here $C$ is the matrix of the charge conjugation which satisfies the conditions
\begin{equation}\label{4a}
C\gamma^{T}_{\alpha} C^{-1}=-\gamma_{\alpha};~~~C^{T}=-C.
\end{equation}
Let us present the field $\psi(x)$ in the form
\begin{equation}\label{5a}
\psi(x)= \frac{1}{\sqrt{2}} \chi_{1} + i  \frac{1}{\sqrt{2}} \chi_{2},
\end{equation}
where
\begin{equation}\label{6a}
\chi_{1}(x)=\frac{\psi(x) + \psi^{c}(x)}{\sqrt{2}};~\chi_{2}(x)=\frac{\psi(x) - \psi^{c}(x)}{\sqrt{2}i}.
\end{equation}
The fields $\chi_{1,2}(x)$ satisfy the Dirac equation
\begin{equation}\label{7a}
i \gamma^{\alpha}\partial_{\alpha} -m)\chi_{1,2}(x)=0
\end{equation}
and {\em additional conditions}
\begin{equation}\label{8a}
\chi^{c}_{1,2}(x)=\chi_{1,2}(x).
\end{equation}
Majorana used the representation in which $\gamma^{\alpha}$ are
imaginary matrices (Majorana representation). In this representation
$\psi^{c}(x)=\psi^{*}(x)$ and $\chi_{1}(x)$ and $\chi_{2}(x)$ are
real and imaginary parts of the field $\psi(x)$. He considered first
the fields $\chi_{1}(x)$ and $\chi_{2}(x)$ and appling 
Jordan-Wigner quantization method he constructed the
quantum field theory of  such fields. Taking into account (\ref{4a}) 
and (\ref{8a})  it is easy to show  that there are no
electromagnetic currents for $\chi_{1,2}(x)$:
\begin{equation}\label{9a}
j_{i}^{\alpha}(x) = \bar\chi_{i}(x)\gamma^{\alpha}\chi_{i}(x)=-\chi_{i}^{T}(x)(\gamma^{\alpha})^{T}\bar\chi_{i}(x)^{T}
=-\bar\chi_{i}(x)\gamma^{\alpha}\chi_{i}(x)=0;~~(i=1,2)
\end{equation}
Therefore,  $\chi_{1,2}(x)$ are fields of particles with electric charge equal to zero.

For the operator  of the energy and momentum  Majorana obtained the
expression
\begin{equation}\label{10a}
P_{i}^{\alpha}=\int \sum_{r}p^{\alpha}a^{\dagger}_{r}(p)a_{
r}(p)d^{3}p ,
\end{equation}
where operators $a_{r}(p)$ and $a^{\dagger}_{r}(p)$ satisfy usual
anticommutation relations.

Thus, $a^{\dagger}_{r}~(p) (a_{ r}(p))$ is the operator of the
creation (absorption) of particle with momentum $p$ and helicity
$r$. There are no states with negative energies and quanta of the
fields $\chi_{1,2}(x)$ are neutral particles which  are identical
to their antiparticles.

Considering complex field $\psi(x)$, presented in the form (\ref{5a}), Majorana came to symmetrical
formulation of the theory of particles and antiparticles with the operators of the total momentum and total charge given by
\begin{equation}\label{11a}
P^{\alpha}=\int \sum_{r}p^{\alpha}[c^{\dagger}_{r}(p)c_{r}(p)+
d^{\dagger}_{r}(p)~d_{r}(p)]  d^{3}p
\end{equation}
\begin{equation}\label{12a}
Q=e\int \sum_{r}[c^{\dagger}_{r}(p)c_{r}(p) -
d^{\dagger}_{r}(p)~d_{r}(p)]  d^{3}p
\end{equation}
Here $c^{\dagger}_{ r}~(p)(c_{ r}(p))$ is the operator of the
creation (absorption) of particle with charge $e$, momentum $p$ and
helicity $r$ and $d^{\dagger}_{ r}~(p)(d_{ r}(p))$ is the operator
of the creation (absorption) of antiparticle with with charge $-e$,
momentum $p$ and helicity $r$. Majorana wrote in the  paper ``A
generalization of Jordan-Wigner quantization method allows not only
to give symmetrical form to the electron-positron theory but also to
construct an essentially new theory for particles without electric
charge (neutrons and hypothetical neutrinos)''. And further in the
paper: ``Although it is perhaps not possible now to ask experiment
to choose between the new theory and that in which the Dirac
equations are simply extended to neutral particles, one should keep
in mind that the new theory is introducing in the unexplored field a
smaller number of hypothetical entities''

Soon after the Majorana paper appeared  Racah \cite{Racah} proposed a method which
could allow to test whether neutrino is Majorana or Dirac particle.
The so-called Racah chain of reactions
\begin{equation}\label{13a}
(A,Z) \to (A,Z+1) + e^{-}+\nu,~~\nu+(A',Z')\to (A',Z'+1) +e^{-}
\end{equation}
is allowed in the case of the Majorana neutrino and is forbidden in
the case of the Dirac neutrino. Of course in 1937 Racah could not
know that even in the case of the Majorana neutrino the chain
(\ref{13a}) is strongly suppressed (see
later).

In 1938 Furry \cite{Furry} for the first time considered neutrinoless double $\beta$-decay of nuclei
\begin{equation}\label{13b}
(A,Z) \to (A,Z+2) + e^{-}+ e^{-}
\end{equation}
induced by Racah chain with virtual
neutrinos. As we will discuss later  the investigation of this process is the major way to
probe the nature of neutrinos.

\section{Neutrino Majorana; basics}

\subsection{Interaction Lagrangian}
Existing weak interaction data are perfectly described by the
Standard Model. The Standard Lagrangians of CC and NC interactions of
neutrinos with other particles are given by
\begin{equation}
\mathcal{L}_{I}^{\mathrm{CC}} = - \frac{g}{2\sqrt{2}} \,
j^{\mathrm{CC}}_{\alpha} \, W^{\alpha} + \mathrm{h.c.};~~
\mathcal{L}_{I}^{\mathrm{NC}} = - \frac{g}{2\cos\theta_{W}} \,
j^{\mathrm{NC}}_{\alpha} \, Z^{\alpha}.\label{1}
\end{equation}
Here $g$ is the $SU(2)$ gauge coupling, $\theta_{W}$ is the weak
angle and
\begin{equation}
j^{\mathrm{CC}}_{\alpha} =2 \sum_{l=e,\mu,\tau}  \bar \nu_{lL}
\gamma_{\alpha}l_{L};~~ j ^{\mathrm{NC}}_{\alpha}
=\sum_{l=e,\mu,\tau} \bar \nu_{lL}\gamma_{\alpha}\nu_{lL}\label{2}
\end{equation}
are charged and neutral currents.

After the discovery of the neutrino oscillations
\cite{SK,SNO,Kamland,K2K,Cl,Gallex,Sage,SKsol}  we know that
neutrino masses are different from zero. Thus, in addition to the
interaction Lagrangian
 (and
kinetic term) {\em a neutrino mass term enter into the total Lagrangian}.

{\em Nature of neutrinos with definite masses are determined by the
type of neutrino mass term.} There are two possible mass terms for
neutrinos, particles with equal to zero electric charges (see
reviews \cite{BilPet,BilGG}): I. Majorana mass term; II.  Dirac mass
term. We will briefly consider first the Dirac mass term.

\subsection{Dirac mass term}

The Dirac mass term has the form
\begin{equation}\label{3}
\mathcal{L}^{\mathrm{D}}=-\sum
_{l',l}\bar\nu_{l'L}\,(M_{\mathrm{D}})_{l'l}\nu_{lL} +\mathrm{h.c.},
\end{equation}
where $M_{\mathrm{D}}$ is a complex $3\times 3$ matrix. After
the standard diagonalization of the matrix $M_{\mathrm{D}}$
for the mass term we have
\begin{equation}\label{4}
\mathcal{L}^{\mathrm{D}} =-\sum^{3}_{i=1}m_{i}\bar \nu_{i}\nu_{i},
\end{equation}
where $\nu_{i}(x)$ is the field of neutrino with mass $m_{i}$.
Flavor fields $\nu_{lL}(x)$ is connected with left-handed fields
$\nu_{iL}(x)$ by the mixing relation
\begin{equation}\label{5}
\nu_{lL}(x)=\sum^{3}_{i=1} U_{li}~\nu_{iL}(x),
\end{equation}
where $U$  is unitary PMNS mixing matrix.

In the case of the Dirac mass term (\ref{3}) the total Lagrangian is
invariant under global gauge transformations
\begin{equation}\label{6}
\nu'_{lL}(x)=e^{i\alpha}~\nu_{lL}(x),~\nu'_{lR}(x)=e^{i\alpha}~\nu_{lR}(x),
~l'(x)=e^{i\alpha}~l(x),~q'(x)=q(x),
\end{equation}
where $\alpha$ is arbitrary constant phase.

Invariance under the transformations (\ref{6}) means that the total
lepton number
\begin{equation}\label{7}
L=L_{e}+L_{\mu}+L_{\tau}
\end{equation}
is conserved and that $\nu_{i}(x)$ is four component field of
neutrinos  ($L(\nu_{i})=1$) and antineutrinos
($L(\bar\nu_{i})=-1$).

\subsection{Majorana mass term}
We will consider now the Majorana mass term. Let us introduce the
column of the left-handed  fields
\begin{eqnarray} n_{L}=\left(
\begin{array}{c}
\nu_{e L}\\
\nu_{\mu L}\\
\nu_{\tau L}\\
\nu_{s_{1} L}\\
\vdots\\
\end{array}
\right).\label{8}
\end{eqnarray}
We assumed that to addition to flavor fields $\nu_{l L}$  
sterile fields $\nu_{s_{i} L}$ can enter into the mass term. From
(\ref{4a}) it follows that
\begin{equation}
(n_{L})^{c}=C~\bar n^{T}_{L} \label{9},
\end{equation}
is the column of right-handed fields.

Majorana mass term is a Lorenz-invariant product of left-handed
components $\bar n_{\alpha' L}$ and right-handed components $
(n_{\alpha L})^{c}$:
\begin{equation}
\mathcal{L}^{\mathrm{M}}=-\frac{1}{2}\, \bar
n_{L}\,M^{\mathrm{M}}(n_{L})^{c} +\mathrm{h.c.}, \label{10}
\end{equation}
where $M^{\mathrm{M}}$ is a symmetrical $(3+n_{s})\times
(3+n_{s})$ matrix ($n_{s}$ is the number of sterile fields).

A symmetrical matrix can be diagonalized with the help of an
unitary matrix:
\begin{equation}
 M^{\mathrm{M}} = U\,m\, U^{T}, \label{11}
\end{equation}
where $U^{\dag}\,U=1$ and $m_{ik}=m_{i}\,\delta_{ik};~~m_{i}>0$.

From (\ref{10}) and (\ref{11}) for the mass term we have
\begin{equation}
\mathcal{L}^{\mathrm{M}}=-\frac{1}{2}~\bar \nu_{L}~m~ (\nu_{L})^{c}
+\mathrm{h.c.},\label{12}
\end{equation}
where
\begin{equation}
\nu_{L}=U^{\dag}~n_{L}.\label{13}
\end{equation}
From (\ref{12}) for the mass term we finally obtain
\begin{equation}
\mathcal{L}^{\mathrm{M}}=-\frac{1}{2}\,\bar\nu ~m ~\nu=
-\frac{1}{2}\,\sum^{3+n_{s}}_{i=1}m_{i}\,\bar
\nu_{i}\nu_{i}.\label{14}
\end{equation}
Here
\bea \nu= \nu_{L}+(\nu_{L})^{c}=\left(
\begin{array}{c}
\nu_{1}\\
\nu_{2}\\
\nu_{3}\\
\vdots\\
\end{array}
\right).\label{15}\eea
We conclude from (\ref{14}) and (\ref{15}) that the field
$\nu_{i}(x)$ is the field of neutrino with mass $m_{i}$ which
satisfy the Majorana condition
\begin{equation}
\nu^{c}_{i}(x)=\nu_{i}(x).\label{16}
\end{equation}
From (\ref{8}) and (\ref{13}) for the neutrino mixing we find
\begin{equation}
\nu_{lL}=\sum_{i=1}^{3+n_{s}}U_{li}\nu_{iL};~~\nu_{s_{k}L}=\sum_{i=1}^{3+n_{s}}U_{s_{k}i}\nu_{iL}.
\label{17}
\end{equation}
The Majorana condition (\ref{16}) is equivalent to the relation
\begin{equation}
\nu_{iR}(x)=(\nu_{iL}(x))^{c}\label{18}
\end{equation}
Therefore, right-handed and left-handed components of the Majorana
fields are connected by the relation (\ref{18}). Let us stress
that in the case of the Dirac field right-handed and left-handed
components are independent.

From Majorana condition   (\ref{16}) for the field $\nu_{i}(x)$ we
obtain
\begin{equation}
\nu_{i}(x)=\int \frac{1}{(2\pi)^{3/2}}\frac{1}{\sqrt{2p^{0}}}~
(e^{-ipx}\,u^{r}(p)\,a^{i}_{r}(p)+e^{ipx}\,v^{r}(p)\,a^{i
\dagger}_{r}(p))\,d^{3}p. \label{19}
\end{equation}
Here  $a^{i}_{r}(p)$ and $a^{i \dagger}_{r}(p)$ are operators of
absorption and creation of neutrino with momentum $p$, helicity $r$
and mass $m_{i}$. Thus,
 {\em Majorana neutrinos and
antineutrinos are identical  particles.}

There exist at present strong arguments
 in favor of Majorana nature
of massive neutrinos. These arguments are based on the fact that
neutrino masses are much smaller than masses of quarks and leptons.
Absolute values of neutrino masses at present are unknown. From the
data of the tritium experiments it was found \cite{Mainz}
\begin{equation}\label{20}
m_{i}\leq 2.3 ~ \rm{eV}
\end{equation}
From different analysis of the cosmological data for the sum of
neutrino masses the  bounds in the range
\begin{equation}\label{21}
\sum_{i}m_{i}\leq (0.4-1.7)~ \rm{eV}
\end{equation}
were inferred \cite{Tegmark}. These bounds are many orders of
magnitude smaller than masses of quarks and leptons.

The most natural explanation of the smallness of neutrino masses is
based on the assumption that the total lepton number is violated by
a right-handed Majorana mass term at a large scale (the famous see-saw
mechanism \cite{see-saw} of neutrino mass generation).

Let us assume that the Dirac mass term is generated by the Standard Higgs mechansm. 
We can expect that eigenvalues of the matrix $M_{\mathrm{D}}$ are of the order of quark or lepton masses.
Taking into account that neutrino masses are much smaller than masses of quarks and leptons
we will assume that lepton number violating right-handed Majorana mass term 
\begin{equation}\label{22}
\mathcal{L}^{\mathrm{M_{R}}}=-\sum
_{l',l}\overline{(\nu_{l'R})^{c}}\,(M_{\mathrm{R}})_{l'l} \nu_{lR} +\mathrm{h.c.},
\end{equation}
with eigenvalues of $M_{\mathrm{R}}$ which are much larger than masses of leptons and quarks, 
 is generated by some new mechanism. We assume also that the left-handed Majorana mass term is equal to 
zero. For the Majorana mass metrix we have in this case 
\begin{eqnarray}
M^{\mathrm{M+D}} = \left(
\begin{array}{cc}
0&  M_{\mathrm{D}}^{T}\\
M_{\mathrm{D}}&  M_{R}\\
\end{array}
\right), \label{23}
\end{eqnarray}
The matrix $M^{\mathrm{M+D}}$ can be
presented in block-diagonal form
\begin{eqnarray}
U^{T}\,M^{\mathrm{M+D}}\,U\simeq \left(
\begin{array}{cc}
m_{\nu}& 0\\
0 & M_{R}\\
\end{array}
\right), \label{24}
\end{eqnarray}
where
\begin{equation}\label{25}
m_{\nu}=-M_{\mathrm{D}}\,M^{-1}_{R}\, M^{T}_{\mathrm{D}}
\end{equation}
is the Majorana neutrino mass matrix.

 From (\ref{25}) it follows
that neutrino masses are much smaller than masses of quarks and
leptons. Values of neutrino masses and neutrino mixing angles can be specified only
in the framework of   concrete models.

If see-saw mechanism is responsible for Majorana neutrino mass
generation in this case heavy Majorana particles, see-saw partners
of light Majorana neutrinos, must exist. CP-violating decays of
these particles in the early Universe is  considered as a plausible
source of the barion asymmetry of the Universe (see review \cite{Barasym}).

\subsection{Majorana mixing matrix}($ |m_{\beta\beta}|\leq1 $   eV)

An unitary $n \times n$ matrix $U $ is characterized by
$\frac{n(n-1)}{2}$ angles and $\frac{n(n+1)}{2}$ phases. The matrix
$U$ can be presented in the form
\begin{equation}\label{26}
U= S^{\dagger}(\beta)~U^{0}~S(\alpha)
\end{equation}
 where
\begin{equation}\label{27}
S_{l'l}(\beta)=e^{i\beta_{l}}\,\delta_{l'l};~~
S_{ik}(\alpha)=e^{i\alpha_{i}}\,\delta_{ik}.
\end{equation}
A common phase is unobservable. Thus one  phase in $S(\alpha)$ and  $S(\beta)$ can be put equal to
zero. We will choose $\alpha_{n}=0$.

Let us consider first the Dirac case. For the CC we have
$2 \sum_{l}  \bar l_{L}
\gamma_{\alpha}\nu_{lL}= 2 \sum_{l,i}  \bar l_{L}
\gamma_{\alpha}U_{li}\nu_{iL}$.
Phases of Dirac fields are
unmeasurable  quantities. Thus, phase factors $e^{i\beta_{l}}$ and
$e^{i\alpha_{i}}$ can be included into the fields
$l(x)$ and $\nu_{i}(x)$, respectively, Therefore, the Dirac mixing matrix is given
by
\begin{equation}\label{28}
U^{D} =U^{0}
\end{equation}
This matrix  is characterized by
$\frac{n(n+1)}{2}-(2n-1)=\frac{(n-1)(n-2)}{2}$ physical phases and
$\frac{n(n-1)}{2}$ angles. In $n=3$ case the Dirac mixing matrix is
characterized by three angles and one phase.

In the case of the  Majorana neutrinos only phase factors
$e^{i\beta_{l}}$ can be absorbed by the Dirac  fields
$l(x)$. Majorana mixing matrix has the form \cite{BilHPet,Schechter}
\begin{equation}\label{29}
U^{M} =U^{0}\,S(\alpha)
\end{equation}
It is characterized by $\frac{n(n-1)}{2}$ angles and
$\frac{n(n+1)}{2} -n = \frac{n(n-1)}{2}$ physical phases. In $n=3$
case the Majorana mixing matrix is characterized by three angles and
three phases.

\subsection{Neutrinoless double $\beta$ decay}

The investigation of the neutrinoless double $\beta$ decay
\begin{equation}\label{30}
    (A,Z) \to (A,Z+2)+e^{-} + e^{-}
\end{equation}
of some even-even nuclei is the most sensitive method which could
allow to reveal the Majorana nature of neutrinos with definite
masses. (see reviews \cite{bbreviews}). In this subsection we will briefly discuss this process. We
will start with the following remarks.
\begin{enumerate}
\item
The investigation of neutrino oscillations in vacuum or in matter
does not allow to  reveal the nature of $\nu_{i}$ \cite{BilHPet,Langacker}.
In fact, the probability of the transition $\nu_{l} \to \nu_{l'}$ in
vacuum is given by (see \cite{BilGG})
\begin{equation}\label{31}
{\mathrm P}(\nu_{l} \to \nu_{l'}) =|\,\sum_{i}U_{l' i} \,~ e^{- i
\Delta m^2_{1i}~ \frac {L} {2E}} ~U^{*}_{li}~ |^{2},
\end{equation}
where $L$ is the distance between neutrino production and neutrino
detection points, $E$ is neutrino energy and $\Delta m^2_{1i}=
m^2_{i}-m^2_{1}$. From (\ref{29})  and  (\ref{31}) it is obvious that additional
Majorana phases $\alpha_{i}$ drop out from the expression for the
transition probability. Thus, we have
\begin{equation}\label{32}
{\mathrm P}^{M}(\nu_{l} \to \nu_{l'})={\mathrm P}^{D}(\nu_{l} \to
\nu_{l'})
\end{equation}
Similarly it can be  shown \cite{Langacker} that the study of  neutrino
transitions in matter also does not allow to reveal the nature of
massive neutrinos.
\item
For the SM weak interaction, theories with massless Dirac and
Majorana neutrinos are equivalent \cite{Okubo}.

We have stressed  before that the major difference between Dirac and
Majorana fields is connected with right-handed components: in the
Dirac case right-handed and left-handed components are independent,
while in the Majorana case right-handed and left-handed components
are connected by the relations (\ref{18}). If $m_{i}= 0$, the
right-handed fields do not enter into the Lagrangian. Hence, there
is no possibility  to distinguish Dirac and Majorana neutrinos in
this case.

For illustration of the equivalence theorem let us consider Racah
chain Eq.(\ref{13a}). From (\ref{1}) and (\ref{2}) it is obvious that
in the first reaction  together $e^{-}$ right-handed  neutrino is
produced. However, in order to produce $e^{-}$ in the second
reaction of the chain left-handed neutrino must be absorbed. Thus,
for massless neutrino the Racah chain is forbidden: neutrino
helicity plays a role of the lepton number.
\item Let us consider the Racah chain in the case of neutrinos with
different from zero masses. In the matrix elements of the process of
production of neutrino with momentum $p$, mass $m_{i}$ and helicity
$r$ enter the spinor $ \frac{1-\gamma_{5}}{2}v^{r}(p)$. Taking into
account linear in $\frac{m_{i}}{2E}$ terms,  we have
\begin{equation}\label{33}
\frac{1-\gamma_{5}}{2}~v^{r}(p)= \frac{1+r}{2}~v^{r}(p)+
r\frac{m_{i}}{2E}\gamma^{0}v^{r}(p),
\end{equation}
where $E$ is neutrino energy.
In the matrix elements of the process of absorption of neutrino with
momentum $p$, mass $m_{i}$ and helicity $r$ enter the spinor
\begin{equation}\label{34}
\frac{1-\gamma_{5}}{2}~u^{r}(p)= \frac{1-r}{2}~u^{r}(p)+
r\frac{m_{i}}{2E}\gamma^{0}u^{r}(p).
\end{equation}
 From (\ref{33}) it follows that in
the neutrino-production process mainly right-handed neutrinos are
produced. From  (\ref{34}) we see that in the cross section of the
absorption of such neutrinos in the second process of the chain
small factors $(\frac{m_{i}}{E})^{2}$ enter. The probability of the
production of the left-handed neutrinos, which have ``large'' weak
absorption cross section, is suppressed by the factor
$(\frac{m_{i}}{E})^{2}$. Therefore, for massive Majorana neutrinos the  Racah chain is suppressed by the
helicity suppression factor $(\frac{m_{3}}{E})^{2}\lesssim
10^{-12}$. (in neutrino processes $E \gtrsim$ MeV). We conclude that it is
not possible in foreseeable future to reveal neutrino nature in
neutrino experiments of the Racah type.
\end{enumerate}
Possibilities to use large targets  (in present-day experiments tens
of kg, in future experiments about 1 ton and may be more), to reach
small background and high energy resolution make experiments on the
search for $0\nu\beta\beta$ decay an unique source of information about the
nature of massive neutrinos $\nu_{i}$
Neutrinoless double $\beta$-decay (\ref{30})
 is
the second order in the Fermi constant process with virtual
neutrinos. For mixed neutrino field
\begin{equation}\label{35}
    \nu_{eL}=\sum_{i}U_{ei}\nu_{iL}
\end{equation}
neutrino propagator is given by
\begin{equation}\label{36}
    \langle 0~|~T(\nu_{eL}(x_{1})~\nu^{T}_{eL}(x_{2}))~|~0 \rangle\simeq
m_{\beta\beta}\,\frac{-i}{(2\,\pi)^{4}} \,\frac{1-\gamma_{5}}{2}\,C~
\int \,e^{-ip(x_{1}-x_{2})}\frac{1}{p^{2}}\,d^{4}p,
\end{equation}
where
\begin{equation}\label{37}
 m_{\beta\beta}= \sum_{i}U^{2}_{ei}\,m_{i}
\end{equation}
is {\em effective Majorana mass}.

For  half-life of $0\nu\beta\beta$-decay the following general
expression can be obtained\cite{bbreviews}
\begin{equation}\label{38}
\frac{1}{T^{0\,\nu}_{1/2}(A,Z)}=
|m_{\beta\beta}|^{2}\,|M(A,Z)|^{2}\,G^{0\,\nu}(E_{0},Z),
\end{equation}
where $M(A,Z) $ is nuclear matrix element and $G^{0\,\nu}(E_{0},Z)$
is known phase-space factor ($E_{0}$ is the energy release).
Nuclear matrix elements are determined only by nuclear
properties and strong interaction and does not depend on neutrino
masses. The calculation of nuclear matrix elements $|M(A,Z)|$ is a
complicated nuclear problem which we will briefly  discuss later on.

There exist at present data of many experiments on the search for $0
\nu \beta\beta$-decay (see \cite{bbdata}).
The stringent lower bound on the half-time of the neutrinoless
double $\beta$ decay was obtained in the germanium Heidelberg-Moscow
experiment\footnote{ An  indication
in favor of $0 \nu \beta\beta$-decay of  $^{76} \rm{Ge}$, found in
\cite{Klop}, is going to be checked by the GERDA experiment started
at Gran Sasso \cite{Gerda}.} \cite{HM}
\begin{equation}\label{39}
T^{0\,\nu}_{1/2}(^{76} \rm{Ge})\geq 1.55 \cdot 10^{25}
\rm{years}~~~(90\% CL)
\end{equation}
Taking into account different calculations of the nuclear matrix
element, from this result for the effective Majorana mass upper
bounds in the following range
\begin{equation}\label{40}
|m_{\beta\beta}| \leq (0.3-1.2)\,~\rm{eV}.
~~~(\rm{Heidelberg-Moscow})
\end{equation}
can be inferred.

The same sensitivity to the effective Majorana mass was reached in
the recent cryogenic experiment CUORICINO \cite{Cuoricino}. For
half-life of $^{130} \rm{Te}$ in this experiment  the following lower bound was found 
\begin{equation}\label{41}
T^{0\,\nu}_{1/2}(^{130} \rm{Te})\geq 1.8 \cdot 10^{24}\,
\rm{years}~~~(90\% CL)
\end{equation}
From this result it was obtained
\begin{equation}\label{42}
|m_{\beta\beta}| \leq (0.2-1.1)\,~\rm{eV}. ~~~(\rm{CUORICINO})
\end{equation}
Several future experiments on the search for  $0\,\nu \beta\,\beta
$-decay (CUORE, MAJORANA, EXO, SUPER-NEMO and others ) are in
preparation at present.\cite{bbfuture}. The aim of these future experiments
is to reach sensitivity
\begin{equation}\label{43}
|m_{\beta\beta}| \simeq \rm{a~few} ~10^{-2} ~\rm{eV}.
\end{equation}
\subsection{The effective Majorana mass}
The observation of $0\nu\beta\beta$-decay would be a direct proof
that massive neutrinos $\nu_{i}$ are Majorana particles. As we will
see in this subsection the determination of the effective Majorana
mass $|m_{\beta\beta}| $ would allow to obtain an important information on the
character of neutrino mass spectrum and the lightest neutrino mass.

All neutrino oscillation data (with the exception of the data of the
LSND experiment \cite{LSND}) are described by the three-neutrino
mixing\footnote{ The LSND indication in favor of $\bar \nu_{\mu}\to
\bar \nu_{e}$ are going to be checked by the  MiniBooNE experiment
\cite{Miniboone}.} The three-neutrino transition probabilities
depend on six parameters: two neutrino mass-squared differences
$\Delta m^{2}_{12}$ and $\Delta m^{2}_{23}$,  three mixing angles
$\theta_{12}$, $\theta_{23}$ and $\theta_{13}$ and CP phase
$\delta$.

From analysis of the Super Kamiokande atmospheric data
the
following 90 \% CL ranges were obtained \cite{SK}
\begin{equation}
1.5\cdot 10^{-3}\leq \Delta m^{2}_{23} \leq 3.4\cdot
10^{-3}\rm{eV}^{2};~~ \sin^{2}2 \theta_{23}> 0.92.\label{44}
\end{equation}
From the global analysis of solar and KamLAND data it was found \cite{SNO}
\begin{equation}
\Delta m^{2}_{12} = 8.0^{+0.6}_{-0.4}~10^{-5}~\rm{eV}^{2};~~~
\tan^{2} \theta_{12}= 0.45^{+0.09}_{-0.07}.\label{45}
\end{equation}

From the result of the reactor CHOOZ experiment \cite {Chooz} for
the parameter $\sin^{2} \theta_{13}$ the following upper bound was
obtained
\begin{equation}\label{46}
\sin^{2} \theta_{13} \leq 5\cdot 10^{-2}
\end{equation}

The current neutrino oscillation experiments do not allow to distinguish
two types of neutrino mass spectra possible in the case of the three-neutrino mixing
\begin{enumerate}
\item Normal spectrum
\begin{equation}
m_{1}<  m_{2}    <  m_{3} ;~ \Delta m^{2}_{12}  \ll    \Delta m^{2}_{23}\label{47}
\end{equation}
\item Inverted spectrum
\begin{equation}
m_{3}<  m_{1}    <  m_{2} ;~ \Delta m^{2}_{12}  \ll
 | \Delta m^{2}_{13}| \label{48}
\end{equation}
\end{enumerate}
For the lightest neutrino mass $m_{0}=m_{1}(m_{3})$ only the upper
bounds (\ref{20}) and (\ref{21}) are  known from the data of the
tritium experiments \cite{Mainz} and cosmological data
\cite{Tegmark}.

Effective Majorana mass $m_{\beta\beta}$ strongly depends on the
value of the lightest neutrino mass and on the type of the neutrino
mass spectrum (see review \cite{Bil} and references therein).
 We will consider three standard neutrino mass spectra.
\begin{enumerate}
\item Hierarchy of neutrino masses
\begin{equation}
 m_{1} \ll m_{2} \ll
m_{3}
\label{51}
\end{equation}
Neglecting the contribution of $m_{1}$,
for the effective Majorana mass we obtain the following expression
\begin{equation}
|m_{\beta\beta}|\simeq\left |\,
 \sin^{2} \theta_{12}\, \sqrt{\Delta m^{2}_{12}} + e^{2i\,\alpha_{23}}
 \sin^{2} \theta_{13}\, \sqrt{\Delta m^{2}_{23}}\,\right|,
 \label{52}
\end{equation}
where $\alpha_{23}=\alpha_{3}- \alpha_{2} $ is the difference of the Majorana CP phases.

The first term in Eq.(\ref{52}) is small because of the smallness of
$\sqrt{\Delta m^{2}_{12}}$. Contribution of the ``large''
$\sqrt{\Delta m^{2}_{23}}$ is suppressed by the small factor
$\sin^{2} \theta_{13} $. From (\ref{44}),  (\ref{45}) and (\ref{46})
for the upper bound of the effective Majorana mass we find
\begin{equation}
|m_{\beta\beta}| \leq 6.6 \cdot 10^{-3}~\rm{eV}.\label{53}
\end{equation}
Thus, upper bound of  $|m_{\beta\beta}|$
in the case of the neutrino mass hierarchy
is smaller that the expected sensitivity of the future  experiments on the search for
$0\nu\beta\beta$-decay. The  value of $|m_{\beta\beta}|$ could be significantly  smaller than (\ref{53})
if cancellation of two terms in  (\ref{52})
takes place.
\item Inverted hierarchy of neutrino masses
\begin{equation}
m_{3} \ll m_{1} <m_{2}. \label{54}
\end{equation}
For the effective Majorana mass we obtain the following expression
\begin{equation}
|m_{\beta\beta}|\simeq \sqrt{| \Delta m^{2}_{\rm{13}}|}\,~ (1-\sin^{2}
2\,\theta_{\rm{12}}\,\sin^{2}\alpha_{12})^{\frac{1}{2}},\label{55}
\end{equation}
where the only unknown parameter  is $\sin^{2}\alpha_{12}$. From (\ref{55}) we have the range
\begin{equation}
\cos  2\,\theta_{\rm{12}} \,\sqrt{ |\Delta m^{2}_{\rm{13}}|}
\leq |m_{\beta\beta}| \leq\sqrt{ |\Delta m^{2}_{\rm{13}}|}
\label{56}
\end{equation}
Taking into account  (\ref{44}) and (\ref{45})   we find that the
effective Majorana mass can take the values
\begin{equation}
0.9\cdot 10^{-2}\leq |m_{\beta\beta}|\leq 5.8\cdot,
10^{-2}~\rm{eV}\label{57}
\end{equation}
which are in the range of
the anticipated sensitivities to $|m_{\beta\beta}|$ of the future
experiments on the search for $0\nu\beta\beta$-decay.
Thus, next generation of the $0\nu\beta\beta$-
experiments can probe the nature of massive neutrinos in the case of the inverted hierarchy of the neutrino
masses.

\item Quasi-degenerate neutrino mass spectrum. If the lightest neutrino mass satisfies inequality
\begin{equation}
m_{0} \gg \sqrt{ |\Delta m^{2}_{\rm{23}}|}
\label{58}
\end{equation}
neutrino mass spectrum is practically degenerate
\begin{equation}
m_{1}\simeq m_{2}\simeq m_{3}
\label{59}
\end{equation}
The effective Majorana mass is given in this case by
\begin{equation}
|m_{\beta\beta}|\simeq m_{0}\, (1-\sin^{2}
2\,\theta_{\rm{12}}\,\sin^{2}\alpha_{12})^{\frac{1}{2}},
\label{60}
\end{equation}
From (\ref{58}) and  (\ref{60}) we conclude that in the case of the
quasi-degenerate spectrum much larger values of the effective
Majorana mass are expected than in the case of the inverted
hierarchy. Such values can be probed in the ongoing
$0\nu\beta\beta$-experiments. Notice that from the observation of
the $0\nu\beta\beta$-decay an information about the value of $m_{0}$
can be inferred:
\begin{equation}
|m_{\beta\beta}|\leq m_{0}\leq  4.4\, |m_{\beta\beta}|
\label{61}
\end{equation}
\end{enumerate}
Three neutrino mass spectra, we have considered,  correspond to different mechanisms of neutrino mass generation (see \cite{Feruglio}).
Masses of quarks and charged leptons satisfy hierarchy of the type  (\ref{51}). Hierarchy of neutrino masses is a typical feature of GUT models (like SO(10)) in which quarks and leptons are unified. Inverted spectrum and quasi-degenerate spectrum require specific symmetries of the neutrino mass matrix.

In order to determine effective Majorana mass from experimental data nuclear matrix elements (NME)
 must be known.  
Two different approaches
are used for the calculation of NME (see reviews  \cite{FSimCivSuh}).:
Nuclear Shell Model and Quasiparticle Random Phase
Approximation . Different calculations of NME differ by factor 2-3
and more. It is important to find a possibility
to test NME calculations.
This will be  possible if $0\nu\beta\beta$-decay of {\em different nuclei} is observed.
If Majorana neutrino mass mechanism is the dominant mechanism of the decay,
 the matrix element of the process is factorized in the form of the product of $m_{\beta\beta}$ and NME.
Thus,  ratio of NME of different nuclei is determined by the ratio of the corresponding half-lives. This can be used as a model independent test of different calculations \cite{BilGri}.

\section{Conclusion}
It was established by the oscillation experiments that neutrino
masses are different from zero and flavor neutrinos $\nu_{e}$,
$\nu_{\mu}$ and $\nu_{\tau}$ are mixed particles. In order to reveal
the origin of the small neutrino masses it is crucial to determine
the nature of neutrinos with definite masses.

There is no theory of neutrino masses at present. There exist
different strategies and models.  One of the most natural (and popular) strategy
is see-saw. The see-saw mechanism is based on the assumption that
the total lepton number $L$ is violated  on a large scale $\simeq
10^{-15}$ GeV by a right-handed Majorana mass term.
If  see-saw mechanism is realized  in this case
neutrinos with definite masses are Majorana particles.

The search for neutrinoless double $\beta$-decay of some even-even
nuclei is the most sensitive method of the investigation of the
nature of neutrinos with definite mass. The sensitivity to the
effective Majorana mass of the future experiments, now under
preparation,  is planned to be about two order of magnitude better
than the sensitivity  of the current
experiments.

{\em  Are massive neutrinos and antineutrinos identical  or
different  ?} This problem, which  has been  put forward by E.
Majorana about 70 years ago, is the most fundamental problem of the
modern neutrino physics. Without its solution the origin of the
small neutrino masses and neutrino mixing can not be revealed.

I would like to acknowledge the Italian program ``Rientro dei
cervelli'' for the support.

\end{document}